\documentstyle[preprint,aps,eqsecnum]{revtex}
\def\mnewpage{\newpage}   

\def\beq#1{\begin{equation} \label{#1}}
\def\eeq{\end{equation}}
\def\bra#1{\left\langle #1\right\vert}
\def\ket#1{\left\vert #1\right\rangle}

\def\NPB{{ Nucl. Phys.} B}
\def\PLB{{ Phys. Lett.} B}

\def\PRD{{ Phys. Rev.} D}

\begin{document}
{
\tighten
\preprint
 {\vbox{\hbox{WIS-96/23/Jun-PH}
 \hbox{TAUP 2343-96}
 \hbox{hep-ph/9606315} }}
 
\title{A Simple General Treatment of Flavor Oscillations}
 
\author{Yuval Grossman\,$^a$ and Harry J. Lipkin\,$^{a,b}$}
 
\footnotetext{Talk given by Harry J. Lipkin at the 18th
annual MRST meeting, Toronto, May 9--10, 1996.}
 
\address{ \vbox{\vskip 0.truecm}
  $^a\;$Department of Particle Physics \\
  Weizmann Institute of Science, Rehovot 76100, Israel \\
\vbox{\vskip 0.truecm}
$^b\;$School of Physics and Astronomy \\
Raymond and Beverly Sackler Faculty of Exact Sciences \\
Tel Aviv University, Tel Aviv, Israel}
 
\maketitle
 
\begin{abstract}%
A unique description avoiding confusion is presented for all flavor-oscillation
experiments in which particles of a definite flavor are emitted from a
localized source. The probability for finding a particle with the wrong flavor
must vanish at the position of the source for all times. This condition
requires flavor-time and flavor-energy factorizations which determine uniquely
the flavor mixture observed at a detector in the oscillation region; i.e.
where the overlaps between the wave packets for different mass eigenstates
differ negligibly from 100\%.
The translation of a ``gedanken'' experiments calculations
(where measurement is perform in time)
is done using the {\it group} velocity.
Energy-momentum (frequency-wave number)
and space-time descriptions are complementary, equally valid and
give the same results.
The two identical phase shifts obtained describe
the same physics; adding them together to get a factor of two is double
counting.
 
\end{abstract}%
 
} 
 
\mnewpage

\section {Introduction}
Flavor oscillations are observed when a source creates a particle
which is a mixture of two or more mass eigenstates, and
a different mixture is observed in a detector. Such oscillations have been
observed in the neutral
kaon and B-meson systems.
In neutrino experiments
it is still unclear whether the eigenstates indeed have
different masses and whether oscillations can be observed.
Considerable confusion has arisen in the description of such
experiments in quantum mechanics \cite{Kayser,NeutHJL},
with questions arising about time dependence and production
reactions \cite{GoldS}, and defining precisely what exactly is
observed in an experiment \cite{Pnonexp}.
Many calculations describe ``gedanken" experiments and require some
recipe for applying the results
to a real experiment \cite{MMNIETO}.
 
We resolve this confusion by noting
and applying one simple general feature of all practical experiments.
The size of the source is small in comparison with the
oscillation wave length to be measured, and
a unique well-defined flavor mixture is emitted by the source; e.g. electron
neutrinos in
a neutrino oscillation experiment. The particles emitted from the source must
therefore be described by a wave packet which satisfies a simple general
boundary condition: the probability amplitude for finding a particle having
the wrong flavor at the source must vanish at all times.
 
This boundary condition requires factorization of the flavor and time dependence
at the position of the source. Since the energy dependence is the Fourier
transform of the time dependence, this factorization also
implies that the flavor dependence of the wave packet is independent of energy
at the position of the source.
In a realistic oscillation experiment the phase is important when the
oscillation length is of the same order as the distance between the
source and the detector.
In that case this
flavor-energy factorization holds over the entire distance between the source
and detector. The boundary condition then determines
the relative phase of components in the wave function with
different mass having the same energy and different momenta. Thus
any flavor oscillations observed as a function of the distance between the
source and detector are described by considering only the interference between
a given set of states having the same energy. All questions of coherence,
relative phases of components in the wave function with different energies and
possible entanglements with other degrees of freedom are thus avoided.
 
Many formulations describe flavor oscillations in time produced by
interference between states with equal momenta and different energies.
These ``gedanken" experiments
have flavor oscillations in time over all space including the source.
We show rigorously that the ratio of
the wave length of the real spatial oscillation to the period of
the gedanken time oscillation is
just the group velocity of the wave packet.
 
\section{Universal Boundary Condition}
 
We now show how the results of a flavor oscillation experiment are completely
determined by the propagation dynamics and the boundary condition that the
probability of observing a particle of the wrong flavor at the position of the
source at any time must vanish. We choose
for example a neutrino oscillation experiment with a source of neutrinos of a
given flavor, say electron neutrinos. The dimensions of the source are
sufficiently small in comparison with the distance to the detector so that it
can be considered a point source
at $\vec x = 0$. The neutrino wave function for this experiment may be a very
complicated wave packet, but a sufficient condition for our analysis is to
require it to describe a pure
$\nu_e$ source at $\vec x = 0$; i.e. the probability of finding a
$\nu_\mu$ or $\nu_\tau$ at $\vec x = 0$ is zero.
 
We first consider propagation in
free space, where the masses and momenta $\vec p_i$ satisfy the usual condition
\beq{WW1b}
{{\vec p}_i\,}^2 = E^2 - m_i^2
\eeq
We expand the neutrino wave function in energy eigenstates
\beq{WW1a}
\psi = \int g(E) dE e^{-iEt}\cdot \sum_{i=1}^3 c_i e^{i\vec p_i\cdot \vec x}
\ket {\nu_i}
\eeq
where $\ket {\nu_i}$ denote the three neutrino mass eigenstates and the
coefficients $c_i$ are energy-independent.
Each energy eigenstate has three terms, one for each mass eigenstate.
In order to avoid spurious flavor oscillations at the source the
particular linear combination of these three terms
required to describe this experiment must be
a pure $\nu_e$ state at $\vec x = 0 $ for each individual energy component.
Thus the coefficients $c_i$ satisfy the conditions
\beq{WW2}
\sum_{i=1}^3 c_i \bra {\nu_i}\nu_\mu \rangle =
\sum_{i=1}^3 c_i \bra {\nu_i}\nu_\tau \rangle = 0
\eeq
The momentum of each of the
three components is determined by the energy and the neutrino masses. The
propagation of this energy eigenstate, the relative phases of its three mass
components  and its flavor mixture at the detector are completely determined by
the energy-momentum kinematics for the three mass eigenstates.
 
The exact form of the energy wave packet described by the function $g(E)$ is
irrelevant at this stage. The components with different energies may be coherent
or incoherent, and they may be ``entangled" with other degrees of freedom of
the system. For example, for the case where a neutrino is produced together with
an electron in a weak decay the function $g(E)$ can also be a function
$g(\vec p_e,E)$ of the electron momentum as well as the neutrino energy.
The neutrino degrees of freedom observed at the detector will then be described
by a density matrix after the electron degrees of freedom have been properly
integrated out, taking into account any measurements on the electron. However,
none of these considerations can introduce a neutrino of the wrong flavor at the
position of the source.
 
Since the momenta $\vec p_i$ are
energy-dependent the factorization does not hold at finite values of $\vec x$.
At very large values of $\vec x$ the wave packet must separate into individual
wave packets with different masses traveling with different velocities
\cite{Nus,Kayser}.
However, for the conditions of a realistic oscillation experiment this
separation has barely begun and the overlap of the wave packets with different
masses is essentially 100\%. Under these conditions the flavor-energy
factorization introduced at the source is still an excellent approximation at
the detector.
 
The flavor mixture at the detector given by substituting the
detector coordinate into Eq. (\ref{WW1a})
can be shown to be the same for all the
energy eigenstates except for completely negligible small differences.
For example, for the case of two neutrinos with energy $E$ and mass eigenstates
$m_1$ and $m_2$ the relative phase of the two neutrino waves at a distance $x$
is:
\beq{WW3a}
\delta \phi(x)= (p_1 - p_2)\cdot x =
{{(p_1^2 - p_2^2)}\over{(p_1 + p_2)}}\cdot x  =
{{\Delta m^2}\over{(p_1 + p_2)}}\cdot x
\eeq
where $\Delta m^2 \equiv m_2^2-m_1^2$.
Since the neutrino mass difference is very small compared to all neutrino
momenta and energies, we use
$|m_2 - m_1| \ll p \equiv (1/2)(p_1 + p_2)$.
Thus we can rewrite eq. (\ref{WW3a})
keeping terms only of first order in $m_2 - m_1$
\beq{WW4a}
\delta \phi(x) =
{{\Delta m^2}\over{2p}}\cdot x =
- \left({{\partial p}\over{\partial (m^2)}}\right)_{E}
\Delta m^2\cdot x
\eeq
where the standard relativistic energy-momentum relation (\ref{WW1b}) gives
the change in energy or momentum with mass when the other is fixed,
\beq{WW4b}
\left({{2E\partial E}\over{\partial (m^2)}}\right)_{p}
= - \left({{2p\partial p}\over{\partial (m^2)}}\right)_{E} = 1
\eeq

Thus we have a complete
solution to the oscillation problem and can give the neutrino flavor as a
function of the distance to the detector by examining the behavior of a single
energy eigenstate. The flavor-energy factorization enables the result to be
obtained without considering any interference effects between different energy
eigenstates. The only information needed
to predict the neutrino oscillations is the behavior
of a linear combination of the three mass eigenstates having
the same energy and different momenta.
All effects of interference or relative phase between
components of the wave function with different energies are time dependent
and are required to vanish at the source, where the flavor is time independent.
This time independence also holds at the detector as long as
there is significant overlap
between the wave packets for different mass states.
The conditions for validity of this overlap condition are discussed below.
 
Neutrino states with the same energy and different momenta are
relevant rather than vice versa
because the measurement is in space, not time, and flavor-time factorization
holds in a definite region in space.
 
\section{Relation between Real and Gedanken Experiments}
We now derive the relation between our result (\ref{WW3a}) which comes from
interference between states with the same energy and different momenta and
the standard treatments using states with the same momentum and different
energies \cite{booknu}.
For the case of two neutrinos with momentum $p$ and mass eigenstates
$m_1$ and $m_2$ the relative phase of the two neutrino waves at a time $t$ is:
\beq{WW6}
\delta \phi(t)= (E_2 - E_1)\cdot t =
\left({{\partial E}\over{\partial (m^2)}}\right)_{p}\cdot
\Delta m^2 \cdot t =
- \left({{\partial p}\over{\partial (m^2)}}\right)_{E}
\Delta m^2 \cdot {{p}\over{E}} \cdot t
\eeq
where we have substituted eq. (\ref{WW4b}).
This is equal to the result (\ref{WW4a})
if we make the commonly used substitution
\beq{WW7b}
x = {{p}\over{E}} \cdot t = v t
\eeq
 
This is now easily generalized to include cases where external fields can
modify the relation (\ref{WW1b}),
but where the mass eigenstates are not mixed. The
extension to propagation in a medium which mixes mass eigenstates e.g. by the
MSW effect \cite{revMSW} is in principle the same, but
more complicated in practice and not considered here.
The relation between energy, momentum and mass is described by an
arbitrary dispersion relation
\beq{WW15}
f(E, p, m^2) = 0
\eeq
where the function $f$ can also be a slowly varying function of the distance
$x$. In that case, the momentum $p$ for fixed $E$ is also a slowly varying
function of $x$. We take this into account by expressing Eq. (\ref{WW4a})
as a differential equation, and defining
the velocity $v$ by the conventional expression for the group
velocity,
\beq{WW16a}
{{\partial^2  \phi(x)}\over{\partial x \partial (m^2)}}=
-\left({{\partial p}\over{\partial (m^2)}}\right)_{E}
= {{1}\over{v}}\cdot \left({{\partial E}\over{\partial (m^2)}}\right)_{p}
\;, \qquad
v \equiv \left({{\partial E}\over{\partial p}}\right)_{(m^2)}
\eeq
Treatments describing real experiments measuring
distances and ``gedanken" experiments measuring time
are seen to be rigorously equivalent if the
group velocity (\ref{WW16a}) relates the two results.
Note that the group velocity and not
the phase velocity enters into this relation.
 
A simple instructive nontrivial example is a toy model dispersion
relation qualitatively
similar to that for a weak gravitational field \cite{Stod,AB}.
A perturbation described by a parameter $\epsilon$ is seen to produce
effects of opposite sign on the oscillation wave length in space
and the period in time
\beq{WW98a}
(1-\epsilon)E^2 - (1+\epsilon){\vec p}^2 =  m^2
\;, \qquad
v = {{(1+\epsilon)p}\over{(1-\epsilon)E}}
\eeq
\beq{WW98b}
\delta \phi(x)
=  {{\Delta m^2}\over{2(1+\epsilon)p}}\cdot x
\;, \qquad
\delta \phi(t) =
{{\Delta m^2}\over{2(1-\epsilon)E}}\cdot t =
{{\Delta m^2}\over{2(1+\epsilon)p}}\cdot v t
\eeq
Errors are avoided by the use of the correct group velocity \cite{AB}.
More realistic examples will be given elsewhere \cite{preparation}.
 
\section{Description in Terms of Time Behavior}
 
It is instructive to describe the same physics in terms of the time
behavior of the wave function.
The specific form of the wave packet given by the function $g(E)$ in
Eq. (\ref{WW1a}) describes the
Fourier transform of the time behavior as seen at
$\vec x = 0 $. This time behavior changes as the packet moves from
the source to the detector. The components corresponding to the different
mass eigenstates move with
different velocities between the source and the detector.
For the case where the wave packets have moved a distance $x$
the centers of the wave packets will have separated by a distance
\beq{WW5a}
\delta x = {{(v_1 - v_2)}\over{v}}\cdot x  \approx
 {{(p_1 - p_2)}\over{p}}\cdot x  =
{{\Delta m^2 }\over{2p^2}}\cdot x
\eeq
where $v_1$, $v_2$ and $v$  denote the individual velocities of the two wave
packets and an average velocity,
and we have assumed that
$m_i^2 = E_i^2 - p_i^2 \ll p_i^2$.
Here it is clearly the group velocity and not the phase velocity which is
relevant, since it is the group velocity which determines the separation
between the wave packets.
This separation between the wave packet centers
produces a phase displacement between the waves at the
detector which is seen to give exactly the
same phase shift as Eq. (\ref{WW3a}).
We see here simply another description of the
same physics, using the complementarity of energy-momentum and space-time
formulations. They are two ways of getting the same answer, not two different
effects that must be added.
 
The same complementarity is seen in the interference between two classical
wave packets moving with slightly different velocities. Even without using
the quantum mechanical relations with energy and momentum there are two possible
descriptions, one using space and time variables and one using frequency and
wave length. The two descriptions are Fourier transforms of one another and
give the same result. Adding the two results is double counting.
 
Such double counting can arise from noting that the particles traveling
with different velocities arrive at the detector at different times
\cite{SWS} and rewriting  Eq. (\ref{WW6}) to obtain
an erroneous factor of 2
which has been extensively discussed\cite{Pnonexp,KaySto,gol,ABMN}.
\beq{QQ9}
x = v_2t_2 = {{p}\over{E_2}}\cdot t_2
= v_1t_1 = {{p}\over{E_1}}\cdot t_1 \;, \qquad
\delta \phi(t) = (E_2t_2 - E_1t_1) = {{\Delta m^2}\over{p}}\cdot x
\eeq
The error arises because it is the centers of the two wave packets that
arrive at different times, not the detected particle.
 
Eventually the wave packet separates into distinct packets,
one for each mass, moving with different velocities. The separation destroys
the flavor-energy and flavor-time factorizations and introduces a time
dependence in the flavor observed at a given large distance.
When the wave
packets for different masses no longer overlap there is no longer any
coherence and there are no further oscillations \cite{Nus}.
The result (\ref{WW3a}) applies for the case where the separation
(\ref{WW5a}) is small compared to the length in space of the wave
packet; i.e. when the eventual separation of the wave packets has barely begun
and can be neglected.
 
\section{Fuzziness in time}
 
The oscillations can be described either in space or in time.
But the distance between the source and the detector is known in a realistic
experiment to much higher accuracy then the time interval.
Thus the interval between the two events of creation and detection has a sharp
distance and a fuzzy time in the laboratory system. The
fuzziness of the time is an essential feature of the experiment
and can be seen explicitly as follows: Let the center of the wave
packet leave the source at $\vec x =0$ at the time $t = 0$.
Then the particle will be observed at the detector at point $x$ at a time
$t \pm \delta t$, where $t$ is the time at which the center of the wave
packet arrives at $x$ and $\delta t$ is the fluctuation in the arrival time
associated with the length of the wave packet. The proper time interval
$\tau$ between emission and detection is given by
\beq{YY901}
\tau^2 = (t \pm \delta t)^2 - x^2 = {{m^2}\over{E^2}} t^2
+ (\delta t)^2 \pm 2t \delta t
\eeq
The length of the wave packet must be sufficiently large
to contain a large number of cycles in order to define a phase,
$E \cdot \delta t \equiv N \gg 1$.
For an oscillation of the order of one cycle to be observed at the time $t$
between two waves differing in energy by $\delta E$ or at the point $x$
between two waves differing in momentum by $\delta p$,
$\delta E \cdot t \approx 1 \approx  \delta p \cdot x$.
Thus
\beq{YY903}
\tau^2 \approx {{m^2}\over{E^2 \delta E^2}} + {{N^2}\over{E^2 }}
 \pm {{2N}\over{E \delta E}}  = {{1}\over{E \delta E}}\cdot \left(
{{m^2}\over{E \delta E}} + {{N^2 \delta E}\over{E }} \pm  2N \right)
\eeq
The uncertainty in the proper time interval due to the finite length of the
wave packet is seen to be much greater than the value of the proper time
interval.
 
The proper time interval between the two events is always fuzzy.
In the laboratory system distance is sharp and time is fuzzy.
A Lorentz transformation to a different frame necessarily mixes distance and
time and makes both fuzzy in a complicated manner. For this reason one must
be careful in interpreting any results obtained in other frames than the
laboratory system.
 
The waves describing the propagation of different mass eigenstates can be
coherent only if the time interval between creation and detection is not
precisely determined and subject to quantum-mechanical fluctuations.
Thus the wave packet created at the source must have a sufficient length in
time (coherence length) to prevent the determination of its velocity with a
precision needed to identify the mass eigenstate.
The small dimensions of the source introduce a momentum uncertainty essential
for the coherence of the waves of different mass eigenstates. The wave packet
describing the experiment must necessarily contain components from different
mass eigenstates with the same energy and different momenta.
 
Conventional experiments
measure distances to a precision with an error tiny in comparison with
the oscillation wave length to be measured. This is easily achieved in the
laboratory. In a ``gedanken" experiment where oscillations in time are
measured, the experimental apparatus must
measure times to a precision with an error tiny in comparison with
the oscillation period to be measured. One might envision an experiment
which measures the time the oscillating particle is created
by observing another particle emitted at the same time; e.g. an
electron emitted in a beta decay together with the neutrino whose oscillation
is observed. But if both the time and position of the created particle are
measured with sufficient precision a very sharp wave packet is created and the
mass eigenstates moving with different velocities quickly separate and
there is no coherence and no oscillation.
 
In reality, when both $x$ and $t$ are measured there are fluctuations in their
values. Using $v=x/t$ the fluctuations in $x$ and $t$ must be large enough to
make the velocity fuzzy. We write $v=v_G \pm \delta v$, where $v_G$ is the
group velocity and $\delta v$ denotes the variation in $v$ due to the
uncertainty in the wave packet.
Then, in order to have oscillation we need
$ \delta v \gg v_{m_1} - v_{m_2}$
where $v_{m_i}$ is the velocity of the $i$'s mass eigenstates.
This is the case in a real experiment.
Typical values are \cite{exper}
$E = O(10\; MeV)$; $x=O(10^2\;m)$; $t=O(10^{-6}\; sec)$
and the relevant masses that can be probed are $\Delta m^2 = O(1\; eV^2)$.
Then, $v_{m_1} - v_{m_2} = O(10^{-12})$.
Since $\delta v \approx dx/x + dt/t$ we
see that the accuracy needed to measure the separate velocities are
$dx = O(10^{-10}\;m)$ and $dt = O(10^{-18}\; sec)$,
far from the ability of present technology.
This calculation can also be performed for all terrestrial experiments, finding
that the present technology is always such that  oscillations can be seen.
 
\section{Conclusions}
 
A unique prescription has been given for the relative phases of the
contributions from different mass eigenstates to a flavor oscillation
experiment with a localized source having a well defined flavor mixture.
The boundary condition that the probability of observing a particle of the
wrong flavor at the source position must vanish for all times requires a
factorization in flavor and energy of the wave function at the position of the
source. This uniquely determines the wave length of the oscillations observed
at the detector as long as the overlap between wave packets for different
mass eigenstates is maintained
at the position of the detector.
 
Whether this wave-packet overlap is sufficiently close to 100\% at the detector
depends upon other parameters in the experiment which determine the detailed
time behavior of the wave packet. If this overlap is appreciable but no longer
nearly complete, the time behavior of the flavor mixture observed at the
detector can be extremely complicated with leading and trailing edges of the
wave packet being pure mass eigenstates and the intermediate region having a
changing flavor mixture depending upon the relative magnitudes of the
contributing mass eigenstates as well as the relative phases.
 
A unique prescription has been given for interpreting results of
calculations for ``gedanken" experiments which measure oscillations in time
for wave packets having the same momentum and different energies. The period
of oscillation in time is related to the wave length of oscillation in space
by the group velocity of the waves.
 
Results are simple in the laboratory system where the positions of the source
and detector are sharp in comparison with all other relevant distances, and
times and proper times must be fuzzy to enable coherent oscillations to be
observed.
 
\section*{Acknowledgments}
We thank Dharam Ahluwalia, Christoph Burgard, Boris Kayser, Pawel Mazur
and Leo Stodolsky for helpful discussions and comments.
One of us (HJL) wishes to thank the Institute of Nuclear Theory at the
University of Washington for its hospitality and to
acknowledge partial support from the U.S. Department of Energy (DOE)
and from the German-Israeli Foundation for Scientific Research and
Development (GIF).
 
{
\tighten

}
 
\end{document}